# Modified Monte Carlo method with thermostat algorithm for model orthonickelates

V. S. Ryumshin[1], Yu. D. Panov[1], V. A. Ulitko[1], A. S. Moskvin[1,2]

[1]Institute of Natural Sciences and Mathematics, Ural Federal University, Ekaterinburg, Russia

[2]Institute of Metal Physics, UD RAS, Ekaterinburg, Russia

E-mail: vitaliy.riumshin@urfu.ru

**Abstract**

The results of numerical simulation using a modified Monte Carlo method with a thermostat algorithm for a pseudospin model of orthonickelates are presented. Temperature phase diagrams are constructed for various degrees of filling and for various parameters of the model, and the effect of local correlations on the critical temperatures of the model orthonickelate is investigated. The possibility of detecting phase inhomogeneous states is shown. The numerical simulation results show good qualitative agreement with the analytical results in the mean field approximation.

# 1 Introduction

Orthonickelates RNiO$_3$ (R is a rare-earth element or Y) exhibit a variety of physical properties, such as metalinsulator transition, unusual conduction behavior, noncollinear magnetic structures, etc. These compounds are of great interest from both a fundamental and an applied side and have been the object of intensive experimental and theoretical research for many years [1–4]. Nickelates belong to a large family of Jahn-Teller magnets, however, the removal of orbital degeneracy in them occurs due to anti- Jahn-Teller disproportionation, which is an experimentally proven fact [5,6]. The same mechanism is associated with the metal-insulator transition, which is also accompanied by a structural phase transition in nickelates, with the exception of LaNiO$_3$, which remains metallic.

Previously, within the framework of the mean field approximation, as well as the classical Monte Carlo (MC) method with kinematic consideration of the constancy of boson concentration, we investigated a two-dimensional analog of the charge-disproportionated phase of rare-earth nickelates RNiO$_3$ with a local basis in the form of a quartet of states consisting of a singlet Ni$^{4+}$ and the triplet Ni$^{2+}$ [7,8]. In this paper, the results of modeling using the thermostat algorithm in the framework of the MC method for a three-dimensional model orthonickelate with a local basis of an octet of states of various charge states of the octahedron NiO$_6$ are discussed. In particular, we study the effect of local charge correlations on the behavior of pure phases, for which only one order parameter is non zero. In addition, as in the mean field approximation and classical MC modeling, there is a tendency to phase separation, as well as the presence of instability regions of the antiferromagnetic insulator and bosonic superfluid phases.

The article is organized as follows. The model used is briefly described in section 2 and some basic results of the mean field approximation are presented. A modified thermostat algorithm of the MC method is described in section 3, and the results of numerical modeling and their comparison with the results of the mean field approximation are presented in section 4. Brief conclusions are formulated in section 5.

## 2 Model and mean field approximation

To describe the electronic structure and phase diagrams of orthonickelates, a generalized model of effective charge triplets was proposed in which the low-energy state of undistorted octahedra $NiO_6$ is formed by a charge triplet $[NiO_6]^{10-,9-,8-}$ (nominally $Ni^{2+,3+,4+}$) with different spin and orbital ground states. The octahedron wave function is denoted by $|\Sigma M; \Gamma\gamma; Sm\rangle$ where $\Sigma M$ – is the charge pseudospin and its projection, $\Gamma\gamma$ is the orbital state ($\Gamma$ - is the irreducible representation of a point group $O_h$ and its row), $Sm$ – is the magnitude and projection of the ordinary spin. State $[NiO_6]^{8-}$ is singlet $|11; A_{1g}1; 00\rangle$, $[NiO_6]^{9-}$ is spin-orbit quartet $|10; E_g0; \frac{1}{2} \pm \frac{1}{2}\rangle$, $|10; E_g2; \frac{1}{2} \pm \frac{1}{2}\rangle$, $[NiO_6]^{10-}$ is spin triplet $|1-1; A_{2g}1; 10\rangle$, $|1-1; A_{2g}1; 1 \pm 1\rangle$.

Let's write down the simplified Hamiltonian of the model leaving only local ($U$) and non-local ($V$) charge correlations, pair transport (t) and bilinear isotropic superexchange ($J$) and limiting the model to interaction with nearest neighbors:

$$\hat{H} = \frac{U}{2}\sum_i \hat{\Sigma}_{z,i}^2 - t \sum_{\langle ij\rangle m}\left(\hat{B}_{m,i}^+\hat{B}_{m,j} + \hat{B}_{m,j}^+\hat{B}_{m,i}\right) + V\sum_{\langle ij\rangle}\hat{\Sigma}_{z,i}\hat{\Sigma}_{z,j} + J\sum_{\langle ij\rangle}\hat{\mathbf{S}}_i\hat{\mathbf{S}}_j. \quad (1)$$

In Hamiltonain (1) $\hat{\Sigma}_z$ is the pseudospin operator, $\hat{\mathbf{S}}$ is the spin operator $S = 1$, and the spin-triplet boson creation and annihilation operators, $\hat{B}_{m,i}^+$ and $\hat{B}_{m,i}$ respectively, are analogous to the creation and annihilation operators in the local (hard-core) boson model [11]. Next, it is convenient to switch to Cartesian components for operators $\hat{B}_{m,i}^+$ and $\hat{B}_{m,i}$ using the relations $\hat{B}_{m,i}^+ = \hat{B}_{xi}^m + i\hat{B}_{yi}^m$, and then the corresponding term will take the form

$$\hat{B}_{m,i}^+\hat{B}_{m,j} + \hat{B}_{m,j}^+\hat{B}_{m,i} = 2\left(\hat{B}_{xi}^m\hat{B}_{xj}^m + \hat{B}_{yi}^m\hat{B}_{yj}^m\right) = 2\hat{\mathbf{B}}_i^m\hat{\mathbf{B}}_j^m. \quad (2)$$

Let's use Bogolyubov's inequality to estimate a large thermodynamic potential:

$$\Omega \leq \tilde{\Omega} = \Psi_0 + \left\langle \hat{H} - \hat{H}_0 - \mu \sum_i \hat{\Sigma}_{z,i} \right\rangle_0, \tag{3}$$

where $\Psi_0$ is the free energy of the ideal system and statistical averaging in the second term is also carried out over the states of the ideal system. The term with a chemical potential $\mu$ allows taking into account the condition of conservation of the number of $e_g$-electrons in the system, which we will write as

$$-\Delta n = \frac{1}{N} \sum_i \langle \hat{\Sigma}_{z,i} \rangle. \tag{4}$$

The value of $\Delta n = 0$ when the lattice is uniformly filled with $e_g$-electrons corresponds to the states of the $Ni^{3+}$, a $\Delta n = -1$ and $+1$ to states $Ni^{4+}$ and $Ni^{2+}$.

We introduce sublattices A and B according to the chessboard type, and choose the Hamiltonian of the ideal system $\hat{H}_0$, in which the interaction of a cluster with its environment is described using molecular fields acting on this cluster in the following form:

$$\hat{H}_0 = \sum_i \hat{H}_{0,i}, \tag{5}$$

$$\hat{H}_{0,i} = \frac{U}{2} \sum_i \hat{\Sigma}_{z,i}^2 - \varphi_{\lambda(i)} \hat{\Sigma}_{zi} - \sum_m f_{\lambda(i)}^m \hat{B}_{m,i} - g_{\lambda(i)} \hat{S}_i. \tag{6}$$

Here the index is $\lambda(i) = A$ or $B$, if the site i belongs to the sublattice $A$ or $B$. The molecular fields $\varphi_\lambda, f_\lambda^m, g_\lambda$ are variational parameters to minimize the estimation of a large thermodynamic potential:

$$\frac{2\tilde{\Omega}}{N} = \sum_\lambda \psi_{0\lambda} - 2zt \sum_m B_A^m B_B^m + zV\Sigma_A \Sigma_B + zJ S_A S_B + \sum_\lambda (\varphi_\lambda - \mu)\Sigma_\lambda \\ + \sum_{\lambda,m} f_\lambda^m B_\lambda^m + \sum_\lambda g_\lambda^m S_\lambda. \tag{7}$$

Here $z$ – is the coordination number (for a simple cubic lattice $z = 6$),

$$\psi_{0\lambda(i)} = -\frac{1}{\beta} \ln(e^{-\beta \hat{H}_{0,i}}), \tag{8}$$

$$\Sigma_\lambda = -\frac{\partial \psi_{0\lambda}}{\partial \varphi_\lambda}, \quad \mathbf{B}_\lambda^m = -\frac{\partial \psi_{0\lambda}}{\partial \mathbf{f}_\lambda^m}, \quad \mathbf{S}_\lambda = -\frac{\partial \psi_{0\lambda}}{\partial \mathbf{g}_\lambda}, \tag{9}$$

where $\beta = 1/(k_B T)$, and we assume $k_B = 1$.

The necessary minimum condition for $\widetilde{\Omega}$ leads to equations for molecular fields:

$$\varphi_\lambda = \mu - zV\Sigma_{\bar\lambda}, \quad \mathbf{f}_\lambda^m = 2zt\mathbf{B}_{\bar\lambda}^m, \quad \mathbf{g}_\lambda = -J\mathbf{S}_{\bar\lambda}, \tag{10}$$

where $\bar\lambda$ is a sublattice additional to $\lambda$. To obtain the dependences of thermodynamic parameters on $\Delta n$ it is necessary to exclude the chemical potential using the equation

$$\Sigma_A + \Sigma_B = -2\Delta n. \tag{11}$$

For the high-temperature disordered (NO) phase, the minimum of $\widetilde{\Omega}$ is reached at $\varphi_\lambda =$ , $\mathbf{f}_\lambda^m =, \mathbf{g}_\lambda = 0$. The condition of loss of stability of the minimum of the NO phase allows finding the critical temperature of the second order phase transition.

Explicit analytical expressions for quantities in the formula (9) can be obtained in some special cases. The charge-ordered (CO) phase with the order parameter $l = (\Sigma_A - \Sigma_B)/2$ is determined by a particular solution of the form

$$\varphi_A \neq \varphi_B, \quad \mathbf{f}_\lambda^m = 0, \quad \mathbf{g}_\lambda = 0. \tag{12}$$

In this case

$$\psi_{0\lambda} = -\frac{1}{\beta}\ln\left[4 + e^{-\beta U/2}\left(e^{\beta\varphi_\lambda} + 3e^{-\beta\varphi_\lambda}\right)\right], \tag{13}$$

$$\Sigma_\lambda(\varphi_\lambda) = \frac{e^{\beta\varphi_\lambda} - 3e^{-\beta\varphi_\lambda}}{4e^{\beta U/2} + e^{\beta\varphi_\lambda} + 3e^{-\beta\varphi_\lambda}}, \tag{14}$$

$$\mathbf{B}_\lambda^m = 0, \quad \mathbf{S}_\lambda = 0. \tag{15}$$

Equation for critical temperature $T_{CO}$

$$zV\Sigma_\lambda'(\varphi) = 1 \tag{16}$$

under the condition $\Sigma_\lambda(\varphi) = -\Delta n$ allows a dependency $T_{CO}(\Delta n)$.

For the antiferromagnetic (AFM) phase with the order parameter $L = \frac{(S_{z,A} - S_{z,B})}{2}$, we consider a particular solution

$$\varphi_\lambda = \varphi, \quad \mathbf{f}_\lambda^m = 0, \quad \mathbf{g}_\lambda = (0,0,g_\lambda). \tag{17}$$

Then $\mathbf{S}_\lambda = (0,0,S_{z,\lambda})$, $\mathbf{B}_\lambda^m = 0$,

$$\psi_{0\lambda} = -\frac{1}{\beta}\ln\{4 + e^{-\beta U/2}[e^{\beta\varphi} + (1 + 2ch(\beta g_\lambda))e^{-\beta\varphi}]\}, \tag{18}$$

$$\Sigma_\lambda(\varphi, g_\lambda) = \frac{e^{\beta\varphi} - [1 + 2ch(\beta g_\lambda)]e^{-\beta\varphi}}{4e^{\beta U/2} + e^{\beta\varphi} - [1 + 2ch(\beta g_\lambda)]e^{-\beta\varphi}}, \tag{19}$$

$$S_{z,\lambda}(\varphi, g_\lambda) = \frac{2sh(\beta g_\lambda)e^{-\beta\varphi}}{4e^{\beta U/2} + e^{\beta\varphi} + [1 + 2ch(\beta g_\lambda)]e^{-\beta\varphi}}. \tag{20}$$

The corresponding equation for the critical temperature $T_{AFM}$ has the form

$$zJ \frac{\partial S_{z,\lambda}(\varphi, 0)}{\partial g_\lambda} = 1. \tag{21}$$

The following solutions can be considered for the phase, which, by analogy with the model of local (hard-core) bosons [11], can be called the phase of a superfluid bose liquid (BS) with the order parameter $\mathbf{B}_\lambda^m \neq 0$

$$\varphi_\lambda = \varphi, \quad \mathbf{f}_\lambda^m = (f^m, 0), \mathbf{g}_\lambda = 0. \tag{22}$$

In this case $\mathbf{S}_\lambda = 0$, $\mathbf{B}_\lambda^m = (B_\lambda^m, 0)$,

$$\psi_{0\lambda} = -\frac{1}{\beta}\ln\{4 + 2e^{-\beta U/2}[e^{-\beta\varphi} + ch(\beta\xi)]\}, \tag{23}$$

where $\xi(\varphi, f) = \sqrt{\varphi^2 + \frac{1}{4}f^2}$, $f^2 = \sum_m(f^m)^2$,

$$\Sigma_\lambda(\varphi, f) = \frac{\varphi sh(\beta\xi) - \xi e^{-\beta\varphi}}{\xi[2e^{\beta U/2} + e^{-\beta\varphi} + ch(\beta\xi)]}, \tag{24}$$

$$B_\lambda^m(\varphi, f) = \frac{f^m sh(\beta\xi)}{4\xi[2e^{\beta U/2} + e^{-\beta\varphi} + ch(\beta\xi)]}. \tag{25}$$

The equation for the critical temperature $T_{BS}$ has the form

$$2zt \frac{\partial B_\lambda^m(\varphi, 0)}{\partial f} = 1. \tag{26}$$

# 3 Monte Carlo thermostat algorithm

We use the thermostat algorithm of the MC method for numerical simulation [12]. Compared with the classical version based on the Metropolis algorithm and assuming continuous change of observables or states at the nodes of the system [7], the quantum nature of the state change in the elementary MC step in our version of the thermostat algorithm eliminates the divergence of heat capacity at low temperatures. The disadvantages of the thermostat algorithm include the high complexity of the elementary MC step, which is associated with the need to solve the eigenvalue problem for the Hamiltonian at the node.

Let us briefly describe the thermostat algorithm for the model (1). Let's write the wave function of the system as the product of the wave functions at the nodes:

$$|\Psi\rangle = \prod_c |\psi_c\rangle, \quad |\psi_c\rangle = \sum_{M\Gamma\gamma Sm} a^c_{M\gamma m}|1M;\Gamma\gamma;Sm\rangle. \quad (27)$$

Let's construct the Hamiltonian of the node c, averaging over the states of all nodes $c' \neq c$:

$$\widehat{H}_c = \langle\widetilde{\Psi}_c|\widehat{H}|\widetilde{\Psi}_c\rangle, \quad |\widetilde{\Psi}_c\rangle = \prod_{c' \neq c}|\psi_{c'}\rangle. \quad (28)$$

For the model (1), we get

$$\widehat{H}_c = \frac{U}{2}\widehat{\Sigma}_z^2 - \mu\widehat{\Sigma}_z - t\sum_{c',m}\left(\langle\widehat{B}_m^+\rangle_{c'}\widehat{B}_m + \langle\widehat{B}_m\rangle_{c'}\widehat{B}_m^+\right) + V\sum_{c'}\langle\widehat{\Sigma}_z\rangle_{c'}\widehat{\Sigma}_z$$
$$+ J\sum_{c'}\langle\widehat{\mathbf{S}}\rangle_{c'}\cdot\widehat{\mathbf{S}}. \quad (29)$$

Here, the values $\langle\widehat{A}\rangle_{c'} = \langle\psi_{c'}|\widehat{A}|\psi_{c'}\rangle$ act as external fields acting on the node c from the node c′, and summation is assumed for the neighbors closest to the node c. A term with a chemical potential has also been added, which allows us to further obtain the dependences of thermodynamic functions on $\Delta n$.

The node state change is implemented using the thermostat algorithm. To do this, we solve the eigenvalue problem on the node:

$$\hat{H}_c|\psi_{c,n}\rangle = \varepsilon_n|\psi_{c,n}\rangle, \tag{30}$$

let us determine the probabilities of the states and construct the distribution function:

$$p_n = \frac{e^{-\beta\varepsilon_n}}{\sum_m e^{-\beta\varepsilon_m}}, \quad F(n) = \sum_{m=1}^{n} p_m, \tag{31}$$

and using a random value evenly distributed over the interval (0,1) $\xi$, we determine the number of the new state of node c from the equation:

$$\xi = F(n). \tag{32}$$

Next, we collect statistics on MC steps in a standard way.

# 4 Comparison of numerical simulation results and mean field approximation

Figure 1 shows the critical transition temperatures from the disordered phase to the CO, AFM, and BS phases of the model nickelate. Here, the parameter of local correlations is $U = 0$, which differs from the situation with $U = -\infty$ discussed earlier in Refs. [7,8]. The simulation results show a characteristic underestimation of $T_{CO}$ compared to the value in the average field method. However, it should be noted that, unlike the classical MC method, there is no reduction in critical temperatures by an order of magnitude. This is probably due to the qualitatively more adequate nature of accounting for temperature fluctuations in the presented algorithm. The discontinuity of the curves for $T_{AFM}$ at $\Delta n \simeq 0.5$ and $T_{BS}$ при $\Delta n \simeq 0.3$ corresponds to the beginning of the instability region of these phases. The ability to define the boundaries of this region is also a feature of this method: certain values of $\Delta n$ are unattainable for all set values of the chemical potential μ. The stability of these phases will be discussed in detail below.

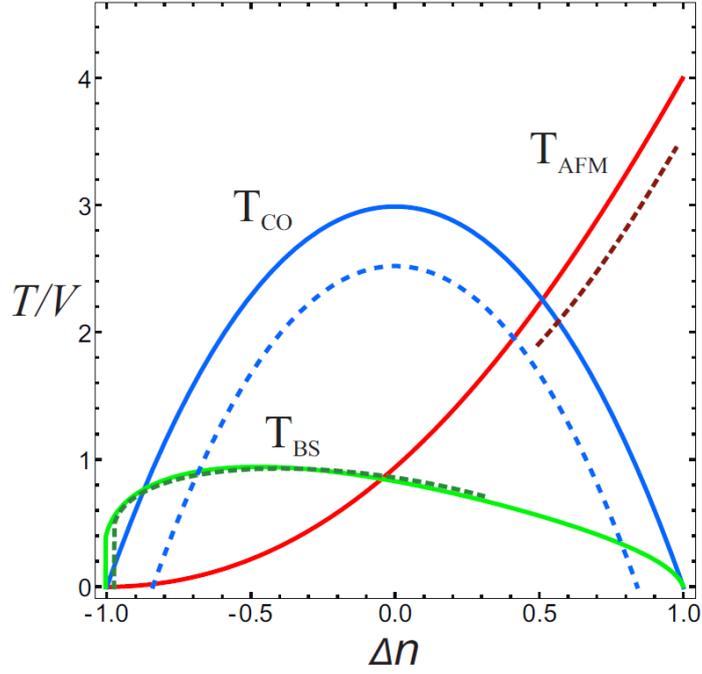

Figure 1. Critical temperature dependencies at z = 6, V = 1, J = 1, t = 1, U = 0. Solid lines — mean field, dotted line — MC simulation.

At $U > 0$ (Figure 2, a) $T_{CO}$ begins to decrease smoothly. A sequential increase of $U$ to 4 (Figure 2, b) causes the appearance of a plateau near the half-fill of $\Delta n = 0$, which indicates the occurrence of a first order phase transition. A further increase of $U$ to 6 (Figure 2, c) leaves only a disordered phase in case of half filling.

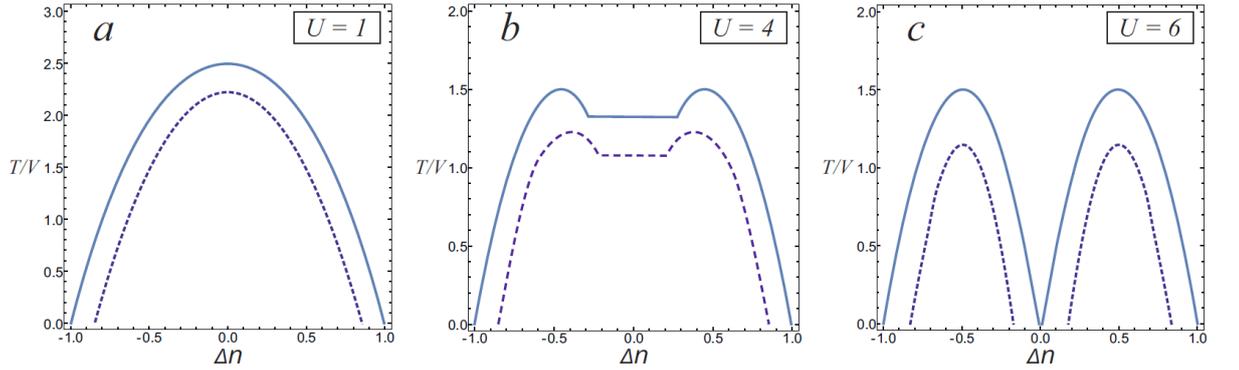

Figure 2. Critical temperature dependencies NO-CO at $V = 1$, $J = 0$, $t = 0$. Solid lines — mean field, dotted line — MC simulation.

Figure 3 shows the dependence of $T_{CO}$ for $\Delta n = 0$ in T-V variables. The results of calculations using the mean field method, equation (16), predict a linear dependence $T_{CO}(V)$, while the MC simulation gives the expected underestimated result. As in the mean field, a second order phase transition is observed at $V/U > 0.5$, and a first order phase transition in the range $0.25 < V/U < 0.4$.

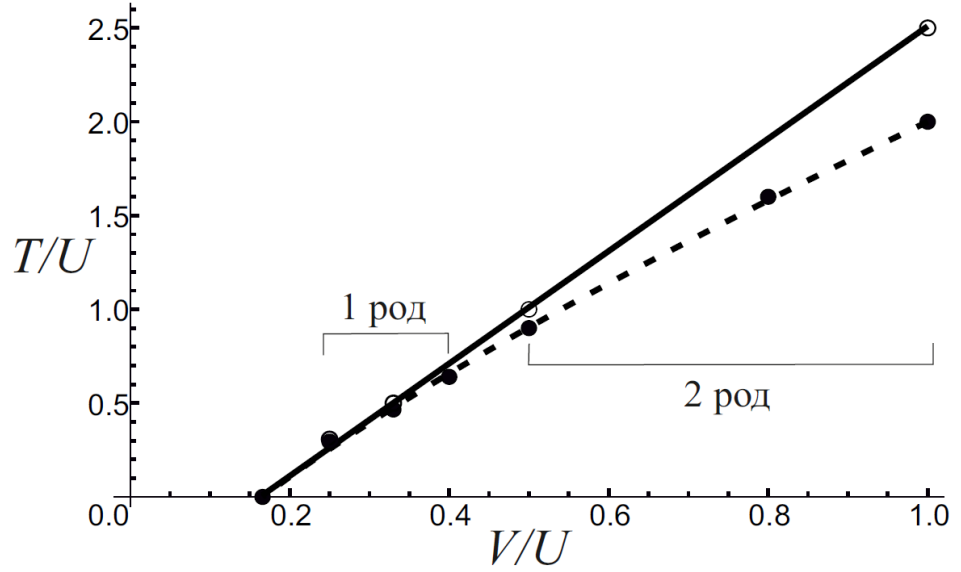

Figure 3. Dependence of the temperature of the NO-CO transition on the parameter of non-local correlations. Solid line — mean field, dotted line — MC simulation.

Figure 4 shows a gradual decrease of TBS with an increase of $U$ to 4, at which a significant suppression of the BS phase occurs. The dotted lines show the boundaries of the thermodynamic stability regions that converge at $\Delta n \simeq 0.6$ with a decrease in temperature for situations with $U = 0, 1$ and 2. In the case of $U = 2$, the BS phase is divided into 2 parts, each of which, in turn, has areas of instability. At $U = 4$, only a small area of the BS phase remains in the range $-1 < \Delta n < 0.7$. Further, the BS phase becomes unstable, which means the appearance of a disordered phase or a region of phase separation. A similar situation is observed for the AFM phase (Figure 5). Its existence is possible in the interval $0.5 < \Delta n < 1$. These results are consistent with data of the mean field approximation and classical MC simulation [7].

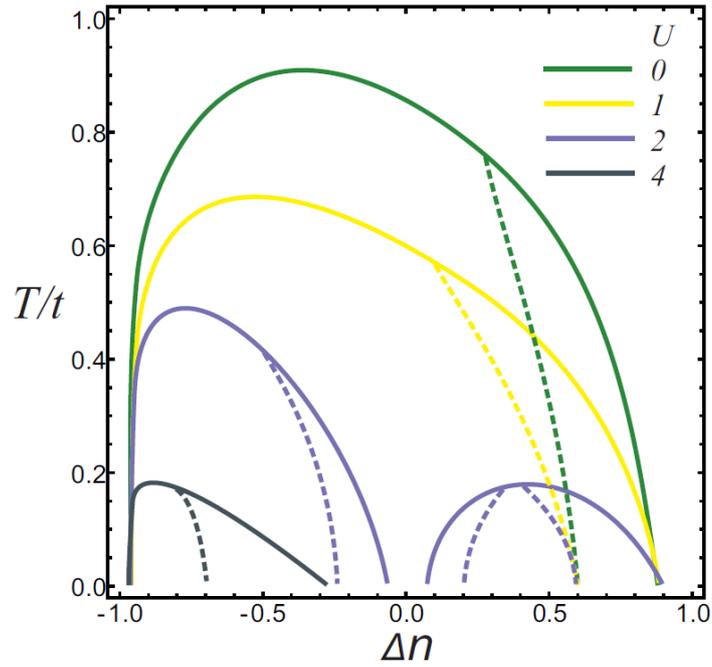

Figure 4. Critical temperature dependencies of NO-BS. The dotted lines show the stability boundaries of the homogeneous phase.

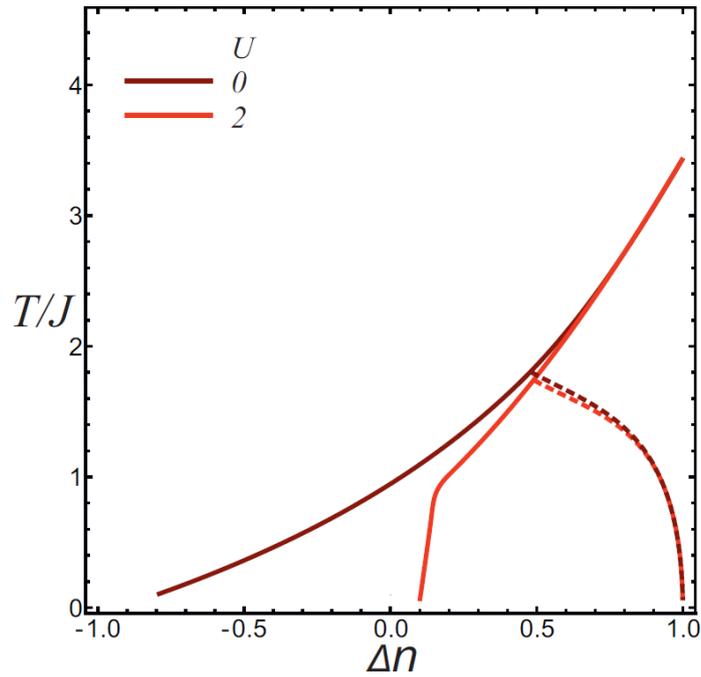

Figure 5. Critical temperature dependencies of NO-AFM. The dotted lines show the stability boundaries of the homogeneous phase.

One of the possible phase diagrams of the model nickelate is shown in Figure 6. Two regions of phase separation can be clearly observed here: CO+BS in the range $-0.6 < \Delta n < 0$ and CO + AFM in the range $0 < \Delta n < 0.9$. The BS, CO, and AFM phases are located sequentially between them. The results show that for the ground state, as well as in the mean-field approximation, the BS and AFM phases are realized near small and large values of n, and the CO phase exists at finite temperatures for these model parameters.

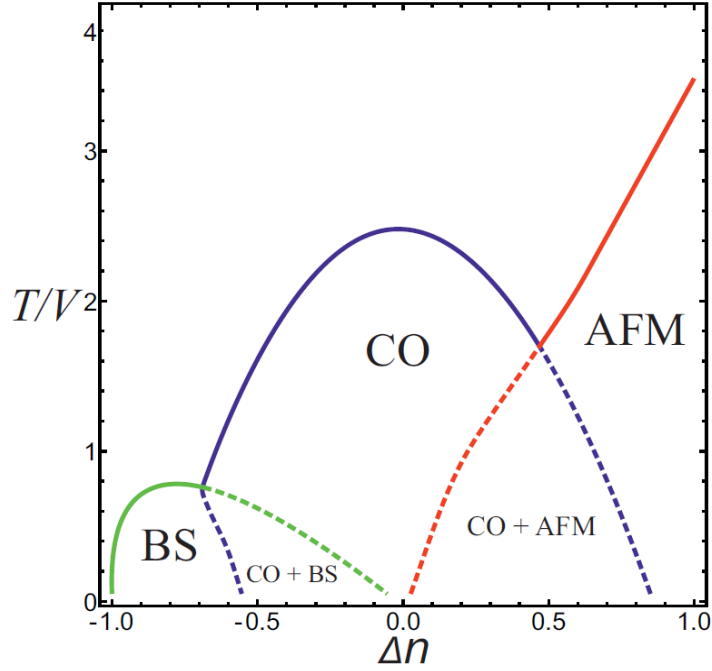

Figure 6. Phase diagram of model nickelate at $V = 1$, $J = 1$, $t = 1$, $U = 0$. Dotted line — boundaries of the CO+BS and CO+AFM phase separation regions.

A comparison of the obtained phase diagrams with half filling $\Delta n = 0.5$ shows that the simulation results correctly predict the NO-CO transition at a qualitative level. However, with a further decrease in temperature, under certain model parameters, it is only possible to preserve the CO phase or transition to the BS phase.

## 5 Conclusion

The results of numerical simulation using the modified Monte Carlo method with the thermostat algorithm show good qualitative agreement with the results of the mean field approximation for those simple solutions when analytical expressions can be obtained. This makes it possible to use this numerical method to analyze the properties of phase states in more complex situations where an analytical solution is not available. Temperature phase diagrams are constructed in this paper for a pseudospin model of rare-earth orthonickelates at various values of the local charge correlations parameter. It is shown that the model predicts various types of phase transitions, as well as the existence of inhomogeneous phase states at different parameter values.

**Funding**

This study was supported by the Ministry of Science and Higher Education of the Russian Federation, project FEUZ-2023-0017.